\documentclass[a4paper,12pt]{article}

\usepackage{graphics}
\usepackage{latexsym}
\newcommand{\cA}{\mathcal{A}}

\newcommand{\cD}{\mathcal{D}}

\newcommand{\cI}{\mathcal{I}}

\newcommand{\cN}{\mathcal{N}}
\newcommand{\cO}{\mathcal{O}}

\newcommand{\cQ}{\mathcal{Q}}
\newcommand{\cS}{\mathcal{S}}

\newcommand{\zetto}{\mathbf{Z}}
\newcommand{\aaru}{\mathbf{R}}

\newcommand{\Qmid}{\cQ_{\mathrm{GRSZ}}}
\newcommand{\tp}{\prime\prime\prime}
\newcommand{\sectiono}[1]{\section{#1}\setcounter{equation}{0}}

\begin{document}
\begin{titlepage}
\thispagestyle{empty}
\begin{flushright}
UT-02-18 \\
hep-th/0204138 \\
April, 2002 
\end{flushright}

\vskip 1.5 cm

\begin{center}
\noindent{\textbf{\LARGE{Comments on Solutions of \vspace{0.5cm}\\ 
Vacuum Superstring Field Theory }}}
\vskip 1.5cm
\noindent{\large{Kazuki Ohmori}\footnote{E-mail: 
ohmori@hep-th.phys.s.u-tokyo.ac.jp}}\\ 
\vspace{1cm}
\noindent{\small{\textit{Department of Physics, Faculty of Science, University of 
Tokyo}} \\ \vspace{2mm}
\small{\textit{Hongo 7-3-1, Bunkyo-ku, Tokyo 113-0033, Japan}}}
\end{center}
\vspace{1cm}
\begin{abstract}
We study classical solutions of vacuum version of Berkovits' superstring field theory, focusing on 
the (super)ghost sector. We first argue that the supersliver state which is annihilated 
by $\eta_0$, though it has the correct quantum numbers and solves the equation of motion, 
is actually non-perturbatively pure-gauge, and hence it fails to describe a D-brane. 
As a step toward the construction of non-trivial solutions, we calculate 
$e^{-\widehat{\Phi}}\widehat{\cQ}e^{\widehat{\Phi}}$ for twisted superslivers. 
As a by-product, we find that the $bc$-twisted sliver solution 
in \textit{bosonic} VSFT can, at least formally, also be written as a pure-gauge configuration. 
\end{abstract}
\end{titlepage}
\newpage
\baselineskip 6mm


\sectiono{Introduction}
In the last year, vacuum string field theory (VSFT)~\cite{RSZ1,VSFT,GRSZ1} attracted much attention. 
The basic idea of this theory has recently been extended to the superstring case 
in~\cite{NSms,ABG,NSgs} for cubic superstring field theory~\cite{CSSFT,ABGKM} and in~\cite{MS} 
for Berkovits' non-polynomial superstring field theory~\cite{SSFT,Brev}. If we assume that the kinetic 
operator $\cQ$ is made purely out of ghost fields as in the bosonic case, the equation of motion admits 
factorized solutions $\Phi=\Phi_{\mathrm{m}}\otimes\Phi_{\mathrm{g}}$, and it was shown in~\cite{MS} 
that we can take $\Phi_{\mathrm{m}}$ to be projectors in the non-polynomial case as well. 

In the (super)ghost sector, however, the construction of brane solutions in Berkovits' theory becomes 
complicated due to the fact that it is formulated in the ``large" Hilbert space,\footnote{The 
$\beta\gamma$-ghost system is fermionized as 
\begin{equation}
\beta=e^{-\phi}\partial\xi, \quad \gamma=\eta e^{\phi}. \label{eq:FMS}
\end{equation}
The usual superstring theory and cubic superstring field theory are closed in the state space 
without the $\xi_0$ mode, which is called the ``small" Hilbert space, whereas the Berkovits' 
superstring field theory uses the full ``large" Hilbert space including $\xi_0$.}
in addition to the existence of the $\beta\gamma$-ghosts and the GSO($\pm$) sectors. Although 
the supersliver state $\Xi$ solves the equation of motion with $\cQ=\Qmid=(c(i)-c(-i))/2i$ and 
hence it was conjectured in~\cite{MS} that the non-BPS D9-brane was described by the 
supersliver itself, we will argue in section~\ref{sec:2} that this conjecture is not correct. 
In particular, any state which is annihilated by $\eta_0$, including the supersliver, is shown to 
be pure-gauge. Such a thing could happen because not only the field configuration space but also 
the space of gauge degrees of freedom has been enlarged in the ``large" Hilbert space formulation. 
\smallskip

To find a clue to the correct D-brane solution, we will pay attention to the fact that there is a 
remarkable resemblance between the ghost number 1 combination $A=e^{-\Phi}\cQ e^{\Phi}$ of the 
string field $\Phi$ in Berkovits' theory and the ghost number 1 string field $\Psi$ of cubic 
string field theory, as explained in section~\ref{sec:4}. Since it has been shown that in bosonic 
vacuum string field theory the equation of motion $\cQ\Psi+\Psi *\Psi=0$ with $\cQ=\Qmid$ is 
solved by the sliver state $\Xi^{\prime}$ in an auxiliary twisted conformal field 
theory~\cite{GRSZ1,HK,Okuyama2,Okuda} and $A$ satisfies an equation of the same form, we 
conjecture that the solution representing the spacetime-filling non-BPS D9-brane is also given by 
$A\simeq\Xi^{\prime}$ with $\cQ=\Qmid$. With this choice the equation of motion $\eta_0A=0$ is satisfied. 
Of course, it is possible that the correct answer will turn 
out to have more complicated structure peculiar to superstring theory: We do not consider such a 
possibility in this paper, except in section~\ref{sec:sum}. Even if we have specified the form of $A$, 
it still remains to express this 
candidate `solution' $A$ in terms of the string field $\Phi$ of this theory. 
Although we have not solved this `inverse' problem yet, we will anyway calculate $e^{-\widehat{\Phi}_0}
\widehat{\cQ}e^{\widehat{\Phi}_0}$ for $\widehat{\Phi}_0=a_-
\Xi^{\tp}\otimes\sigma_1$ to show an example in which $e^{-\widehat{\Phi}}\widehat{\cQ}e^{\widehat{\Phi}}$ 
is exactly calculable. Here, $\Xi^{\tp}$ is the supersliver state in the 
`triply-twisted' conformal field theory introduced in section~\ref{sec:3}, where the twisting 
is performed in all of the $bc$-, $\phi$- and $\eta\xi$-sectors of the ghost CFT. 
The configuration $\widehat{\Phi}_0$ has the following preferable features: $\widehat{\Phi}_0$ 
has ghost number 0 and picture number 0; $\Xi^{\tp}$ lives in the GSO($-$) sector, while $\Xi$ 
is in the GSO($+$) sector; $\widehat{\Phi}_0$ has the factorized form $\Xi_{\mathrm{m}}\otimes
\widehat{\Phi}_{\mathrm{g}}$ and its matter part is simply the matter sliver; $e^{\widehat{\Phi}_0}$ 
is exactly calculable; Neither $\widehat{\eta}_0$ nor $\widehat{\cQ}$ annihilates $\widehat{\Phi}_0$ 
so that it may not be pure-gauge. Therefore, we believe that the triply-twisted supersliver $\Xi^{\tp}$ 
may be useful in constructing a non-trivial solution in non-polynomial vacuum superstring field theory. 

We further calculate $A$ for another configuration $\Phi_e=a_+e^{-\frac{1}{2}
\rho(i)}e^{-\frac{1}{2}\bar{\rho}(i)}|\Xi^{\prime}\rangle$, 
and we will obtain the result $\Xi^{\prime}\propto
e^{-\Phi_e}\cQ e^{\Phi_e}$, where $\Xi^{\prime}$ denotes the $bc$-twisted sliver of ghost number 1. 
In the context of \textit{bosonic} vacuum string field theory, it looks like this equation says that 
the $bc$-twisted sliver, which is the only ghost solution obtained so far, is a pure-gauge 
configuration. We will have a discussion on this point in section~\ref{sec:5}. 

Section~\ref{sec:sum} is devoted to summary of our results and discussion for future directions. 
In Appendices we derive some formulae about the $*$- and $*^{\tp}$-products used in the text.

\sectiono{Problems with the Supersliver as a Brane Solution of Vacuum Superstring Field 
Theory}\label{sec:2}
In this paper we mainly deal with the issues of non-polynomial vacuum superstring field theory, 
which is supposed to describe the open string dynamics around the tachyon vacuum in 
Berkovits' superstring field theory on a non-BPS D9-brane of type IIA theory. 
This type of vacuum superstring field theory was elegantly formulated by Mari{\~n}o and 
Schiappa in~\cite{MS}, based on the earlier work of Kluso\v{n}~\cite{Klu5}. 
We will discuss the ghost structure of non-perturbative brane solutions in this theory. 
\medskip

First of all, we see that the full (\textit{i.e.} matter+ghost) geometric supersliver, 
although it is a solution of the equation of motion, fails to describe the non-BPS D9-brane. 
From now on, we will use the notation 
\begin{equation}
A(\Phi)=e^{-\Phi}\cQ e^{\Phi}, \label{eq:CA}
\end{equation}
where $\Phi$ is the string field of ghost number 0 and picture number 0, and $\cQ$ is the 
kinetic operator of vacuum superstring field theory. From the Wess-Zumino-Witten--like 
action\footnote{When discussing the tachyon condensation, we need to include in the theory the 
GSO($-$) sector where the tachyon lives. It can be achieved by introducing the internal Chan-Paton 
factors~\cite{TPNS,BSZ} as $\widehat{\Phi}=\Phi_+\otimes\mathbf{1}+\Phi_-\otimes\sigma_1$, where 
$\sigma_1$ is one of the Pauli matrices and $\Phi_+,\Phi_-$ are the string fields in the 
GSO($+$), GSO($-$) sectors, respectively. In this section, however, since the supersliver state 
is fully contained in the GSO($+$) sector, we will forget about this 
structure and simply write $\Phi\equiv\Phi_+$ for a while.}~\cite{SSFT}
\begin{equation}
S=\frac{\kappa_0}{2}\int\left(\left(e^{-\Phi}\cQ e^{\Phi}\right)\left(e^{-\Phi}\eta_0 e^{\Phi}
\right)-\int\limits_0^1dt\left(e^{-t\Phi}\partial_te^{t\Phi}\right)\left\{e^{-t\Phi}\cQ e^{t\Phi},
e^{-t\Phi}\eta_0e^{t\Phi}\right\}\right), \label{eq:WZW}
\end{equation}
we derive the following equation of motion:
\begin{equation}
\eta_0A=\eta_0\left( e^{-\Phi}\cQ e^{\Phi}\right)=0. \label{eq:CB}
\end{equation}
(For more details about Berkovits' superstring field theory, see~\cite{SSFT,BSZ,Brev,0102085,Smet}.)
Let us define the (geometric) supersliver state $\Xi$ as 
\begin{equation}
\langle\Xi |\varphi\rangle =\langle f\circ\varphi(0)\rangle_{\mathrm{UHP}}, \label{eq:CC}
\end{equation}
where $f(\xi)=\tan^{-1}\xi$ and $|\varphi\rangle$ is an arbitrary Fock space state. 
Defined this way, $\Xi$ can be shown to be a projector of the $*$-algebra, $\Xi *\Xi =\Xi$, 
in the same way as in the bosonic case~\cite{RSZ4}. (The algebraic construction 
of the (twisted) sliver state was given 
in~\cite{KP,RSZ2}\cite{David,HK,HM,Kishimoto,Okuyama1}\cite{NSms, MS}\cite{NSgs}.) 
Note that the sliver state, as well as any surface state $|\Sigma\rangle$,
is annihilated by $\eta_0$:
\begin{equation}
\langle\Sigma |\eta_0|\varphi\rangle=\left\langle f_{\Sigma}\circ\left(\oint\limits_C\frac{dz}{2\pi i}
\eta(z)\cdot\varphi(0)\right)\right\rangle_{\mathrm{UHP}}=\left\langle\oint\limits_{C^{\prime}}
\frac{dz}{2\pi i}\eta(z)\cdot f_{\Sigma}\circ\varphi(0)\right\rangle_{\mathrm{UHP}}=0, 
\label{eq:CG}
\end{equation}
where $f_{\Sigma}(\xi)$ is the conformal map associated with the Riemann surface $\Sigma$, 
and $C,C^{\prime}$ are the integration contours encircling $0,f_{\Sigma}(0)$, respectively. 
The second equality holds because $\eta_0$ is the contour integral of a primary field of 
conformal weight 1. The last equality can be shown by the contour-deformation argument. 
More strongly, it can be shown that $\langle\Sigma|\eta_0|\cA\rangle$ vanishes for 
arbitrary states $|\cA\rangle$ even if they do not belong to the Fock space. 
Any surface state $|\Sigma\rangle$ can be written as~\cite{LPP,RZ,Wedge} 
\begin{equation}
|\Sigma\rangle=U^{\dagger}_{f_{\Sigma}}|0\rangle \label{eq:XA}
\end{equation}
with
\begin{equation}
U^{\dagger}_{f_{\Sigma}}=e^{\sum v_n L_n}, \label{eq:XB}
\end{equation}
where $v_n$'s are coefficients determined by the conformal map $f_{\Sigma}$, and 
$L_n$'s are the total Virasoro generators. Since the superghost sector of $L_n$ 
can be expressed in terms of $\beta\gamma$ ghosts themselves, $L_n$ does not contain 
the zero mode of $\xi$, as can be seen from the fermionization formula~(\ref{eq:FMS}). 
In addition, the $SL(2,\aaru)$ vacuum $|0\rangle$ is annihilated by $\eta_0$ because $\eta$ 
has conformal weight 1. Hence $|\Sigma\rangle$ does not give rise to any factor of $\xi_0$. 
If we recall that the correlation function in the ``large" Hilbert space is normalized as 
\begin{equation}
\langle 0|\xi_0 c_1c_0c_{-1}e^{-2\phi}|0\rangle=V_{10}, 
\end{equation}
where $V_{10}$ denotes the volume of the 10-dimensional space-time, $|\cA\rangle$ must 
supply two factors of $\xi_0$ in order for $\langle\Sigma|\eta_0|\cA\rangle$ to have 
a non-vanishing value, but it is impossible. We then conclude that $\eta_0|\Sigma\rangle$ strongly 
vanishes in the sense that it has vanishing inner product with any state. 

Now consider a string field $\Phi_a$ which is annihilated by $\eta_0$, such as the sliver. 
From the derivation property of $\eta_0$, it follows 
\begin{equation}
\eta_0 e^{\Phi_a}=\eta_0 e^{-\Phi_a}=0. \label{eq:RA}
\end{equation}
Given that $\{\eta_0,\cQ\}=0$, we find that $\Phi_a$ trivially satisfies the equation 
of motion~(\ref{eq:CB}) as 
\begin{equation}
\eta_0\left(e^{-\Phi_a}\cQ e^{\Phi_a}\right)=(\eta_0 e^{-\Phi_a})(\cQ e^{\Phi_a})-e^{-\Phi_a}
\cQ (\eta_0 e^{\Phi_a})=0. \label{eq:RB}
\end{equation}
Since the equation of motion does not constrain the form of $\Phi_a$ anymore, we cannot 
hope to obtain any isolated solutions even after fixing the gauge. 
This is obviously an unwelcome feature when looking for the D-brane solutions. 
Moreover, the energy density associated with such a state is completely zero, as can be 
seen from the form of the action~(\ref{eq:WZW}). This problem cannot be resolved even if 
we take the overall normalization factor $\kappa_0$ to be infinite, because there are no sources 
of $\xi_0$ which is necessary to have non-vanishing correlators in the ``large" Hilbert space 
formulation. In fact, we will show below that 
any such state $\Phi_a$ is pure-gauge in Berkovits' superstring field theory. 
\smallskip

In what follows, we consider the action~(\ref{eq:WZW}), but we do not restrict $\cQ$ to being pure 
ghost, if only $\cQ$ satisfies the requisite properties: nilpotency, Leibniz rule, etc. 
By analogy with the Wess-Zumino-Witten action for a field $g=e^{\Phi}$ which takes value in a 
group manifold $G$, we see that the action~(\ref{eq:WZW}) is invariant under the finite gauge transformation 
\begin{equation}
e^{\Phi}\quad\longrightarrow\quad (e^{\Phi})^{\prime}=h_1^{\cQ}*e^{\Phi}*h_2^{\eta_0}, \label{eq:CH}
\end{equation}
where the superscripts indicate that the gauge parameters 
$h_1^{\cQ},h_2^{\eta_0}$ are annihilated by $\cQ,\eta_0$, respectively. 
Then we can take $h_2^{\eta_0}=e^{-\Phi_a}$ because of~(\ref{eq:RA}). 
By choosing $h_1^{\cQ}$ to be the identity string field $\cI$,\footnote{For 
the identity string field defined geometrically as $\langle\cI |
\varphi\rangle=\langle f_{\cI}\circ\varphi(0)\rangle_{\mathrm{UHP}}$ with $f_{\cI}(\xi)=2\xi /(1-\xi^2)$, 
we can show $\cI *A=A*\cI=A$ using the gluing theorem~\cite{KiOh}. (\textit{cf.} \cite{Wedge}) 
On the other hand, the oscillator representation of the identity state~\cite{GJ} does \textit{not} 
satisfy $\cI *A=A*\cI=A$ in the ghost sector~\cite{Kishimoto}. We take the identity string field to be 
the geometric one in the present paper, because it is important that the identity $\cI$ behaves as the 
`identity element' of the $*$-algebra.} 
we find from~(\ref{eq:CH}) 
\[ e^{\Phi_a}\quad\longrightarrow\quad \cI *e^{\Phi_a}*e^{-\Phi_a}=e^0, \]
\textit{i.e.} $\Phi_a$ has been gauged away. This completes the proof 
that the string field obeying $\eta_0\Phi=0$ is non-perturbatively pure-gauge. 
This fact can more transparently be seen in the case of a projector. When $h_1^{\cQ}=e^{\cQ\Omega}$ 
and $h_2^{\eta_0}=e^{\eta_0\Lambda}$ with $\Omega,\Lambda$ infinitesimal gauge parameters, the infinitesimal 
form of~(\ref{eq:CH}) is, as is well-known, given by 
\begin{equation}
\delta (e^{\Phi})=(\cQ\Omega) *e^{\Phi}+e^{\Phi}*(\eta_0\Lambda). \label{eq:CJ}
\end{equation}
Given that $\Phi *\Phi=\Phi$ and $\eta_0\Phi=0$, by taking $\Omega=0$ and $\Lambda=a\xi_0\Phi$ 
$(a\ll 1)$ we have
\[ (e-1)\delta\Phi=\left(\cI+(e-1)\Phi\right) *\eta_0(a\xi_0\Phi)=ae\Phi\propto\Phi, \]
which implies that $\Phi$ can be gauged away by piling up the above infinitesimal transformations. 
The conclusion that any string field configuration annihilated by $\eta_0$ is pure gauge seems very 
restrictive from the viewpoint of vacuum superstring field theory because the $*$-algebra projectors 
known so far~\cite{GRSZ3,Schnabl} are constructed within the ``small"  Hilbert space and all annihilated 
by $\eta_0$. For the same reason, the configurations $\Phi$ which are annihilated by $\cQ$ are 
also pure gauge, but this criterion crucially depends on the choice of $\cQ$. 
\medskip

We remark here that our concerns above are irrelevant to 
cubic superstring field theory~\cite{CSSFT,ABGKM,NSms,ABG,NSgs} 
which is formulated within the ``small" Hilbert space. In that case it has been proposed that 
the space-filling non-BPS D9-brane is represented by the sliver states in the 
twisted conformal field theories~\cite{NSgs}.

\sectiono{Triply Twisted Conformal Field Theory}\label{sec:3}
In the preceding section, we have seen that the supersliver, though it solves the equation of motion 
and has the correct quantum numbers (ghost number 0 and picture number 0), is not qualified as a 
D-brane solution. To obtain a non-trivial solution by deforming the 
sliver,\footnote{Throughout this and the next sections the `sliver' is interchangeable with any one of the 
projectors such as the butterfly state and the nothing state~\cite{GRSZ3,Schnabl}.}
it will be useful to introduce an auxiliary twisted conformal field theory, according 
to~\cite{GRSZ1}. We twist the energy-momentum tensors $T,\overline{T}$ by the ghost and picture number 
currents $j^{bc}=-bc,j^{\phi}=-\partial\phi,j^{\eta\xi}=-\eta\xi;\ \overline{j^{bc}}=-\bar{b}\bar{c},
\overline{j^{\phi}}=-\bar{\partial}\bar{\phi},\overline{j^{\eta\xi}}=-\bar{\eta}\bar{\xi}$ 
as\footnote{Similar operation in $\phi$-sector has been considered in~\cite{NSgs}.}
\begin{eqnarray}
T^{\tp}(w)&=&T(w)-\partial j^{bc}(w)-\partial j^{\phi}(w)-\partial j^{\eta\xi}(w), \label{eq:DA} \\
\overline{T}^{\tp}(\bar{w})&=&
\overline{T}(\bar{w})-\bar{\partial} \overline{j^{bc}}(\bar{w})-\bar{\partial} \overline{j^{\phi}}
(\bar{w})-\bar{\partial}\overline{j^{\eta\xi}}(\bar{w}), \nonumber
\end{eqnarray}
such that $b^{\prime},c^{\prime},\xi^{\prime},\eta^{\prime}$ in the new theory have spins 1, 0, 1, 0, 
respectively. 
This new `triply-twisted' conformal field theory, denoted by CFT$^{\tp}$, has a total central 
charge $c^{\tp}=15-2-1-2=10$. If we bosonize the $bc$- and $\eta\xi$-systems 
as\footnote{Hermitian bosonized 
ghost fields $\rho,\bar{\rho}$ are related to the antihermitian fields $\varphi,\bar{\varphi}$ 
used in~\cite{GRSZ1} via $\rho=i\varphi,\bar{\rho}=i\bar{\varphi}$.} 
\begin{eqnarray}
& &c\simeq e^{\rho},\ b\simeq e^{-\rho} ,\ \xi\simeq e^{\chi},\ \eta\simeq e^{-\chi}, \label{eq:DB} \\
& &\bar{c}\simeq e^{\bar{\rho}},\ \bar{b}\simeq e^{-\bar{\rho}}, \bar{\xi}\simeq e^{\bar{\chi}},\ 
\bar{\eta}\simeq e^{-\bar{\chi}}; \nonumber \\
& &c^{\prime}\simeq e^{\rho},\ b^{\prime}\simeq e^{-\rho}, \xi^{\prime}\simeq e^{\chi},\ 
\eta^{\prime}\simeq e^{-\chi}, \label{eq:DC} \\
& & \bar{c}^{\prime}\simeq e^{\bar{\rho}},\ \bar{b}^{\prime}\simeq e^{-\bar{\rho}}, \bar{\xi}^{\prime}
\simeq e^{\bar{\chi}},\ \bar{\eta}^{\prime}\simeq e^{-\bar{\chi}}; \nonumber 
\end{eqnarray}
the relation between the Euclidean world-sheet actions $\cS$ and $\cS^{\tp}$ of the two theories 
is found to be 
\begin{equation}
\cS^{\tp}=\cS-\frac{1}{2\pi}\int d^2w \sqrt{g}R(g)\left(\rho+\bar{\rho}-\phi-\bar{\phi}+\chi+\bar{\chi}\right), 
\label{eq:DD}
\end{equation}
where $w=\sigma+i\tau$ 
denotes the world-sheet coordinates, $g$ and $R(g)$ are the world-sheet metric and its scalar 
curvature, respectively. The OPEs among the bosonized ghosts are 
given by 
\begin{eqnarray}
& &\rho(z)\rho(0)\sim +\log z, \quad \phi(z)\phi(0)\sim -\log z, \quad \chi(z)\chi(0)\sim +\log z, \nonumber \\
& &\bar{\rho}(\bar{z})\bar{\rho}(0)\sim +\log \bar{z}, \quad \bar{\phi}(\bar{z})\bar{\phi}(0)
\sim -\log \bar{z}, \quad \bar{\chi}(\bar{z})\bar{\chi}(0)\sim +\log \bar{z}, \label{eq:DE} \\
& &0\qquad \mathrm{otherwise}.  \nonumber 
\end{eqnarray}
The conformal weights of the ghost fields are summarized as follows:
\begin{equation}
	\begin{array}{|c|c|c|c|}
	\hline
	 & e^{\ell \rho} & e^{\ell\phi} & e^{\ell\chi} \\ \hline 
	\mbox{in the original theory} & (\frac{1}{2}\ell (\ell-3),0) & 
	(-\frac{1}{2}\ell(\ell+2),0) & (\frac{1}{2}\ell(\ell-1),0) \\ \hline
	\mbox{in the twisted theory} & (\frac{1}{2}\ell (\ell-1),0) & 
	(-\ell^2/2,0) & (\frac{1}{2}\ell(\ell+1),0) \\ \hline 
	\end{array} \label{eq:DF}
\end{equation}
and similarly for antiholomorphic fields. 
On the flat world-sheet (\textit{e.g.} each local coordinate patch) 
$\rho,\phi,\chi$ fields of the two theories can be identified:
\[ e^{\ell\rho(z)}\Big|_{\mathrm{CFT}}=z^{d^{\tp}-d}e^{\ell\rho(z)}\Big|_{\mathrm{CFT}^{\tp}}
\quad \mathrm{etc.}, \]
where $d$ and $d^{\tp}$ denote the conformal weights~(\ref{eq:DF}) of $e^{\ell\rho}$ in the 
respective theories. Therefore we have the following identification among the oscillators of the 
two theories, 
\begin{equation}
c_n\leftrightarrow c_n^{\prime},\quad b_n\leftrightarrow b_n^{\prime},\quad \xi_n\leftrightarrow
\xi_n^{\prime},\quad \eta_n\leftrightarrow\eta_n^{\prime}. \label{eq:TH}
\end{equation}
The $SL(2,\aaru)$-invariant 
vacua $|0\rangle$ and $|0^{\tp}\rangle$ of the two theories are mapped as 
\begin{equation}
\xi ce^{-\phi}(0)|0\rangle \ \ \longleftrightarrow \ \ |0^{\tp}\rangle. \label{eq:DG}
\end{equation}
For any state $|\varphi\rangle$ we can give vertex operator representations of it with respect 
to the $SL(2,\aaru)$ vacua of the two theories as 
\begin{equation}
|\varphi\rangle=\varphi(0)|0\rangle=\varphi^{\tp}(0)|0^{\tp}\rangle. \label{eq:DH}
\end{equation}
The correlation functions of the two theories are normalized as 
\begin{equation}
\left\langle\xi\frac{1}{2}c\partial c\partial^2ce^{-2\phi}\right\rangle_{\mathrm{UHP}}
=\langle\eta^{\prime}c^{\prime}\rangle^{\tp}_{\mathrm{UHP}}=V_{10}. \label{eq:DI}
\end{equation}

The star product of the twisted theory is defined as 
\begin{equation}
\langle\Phi_1|\Phi_2*^{\tp}\Phi_3\rangle=\left\langle f_1^{(3)}\circ\Phi_1^{\tp}(0)f_2^{(3)}
\circ\Phi_2^{\tp}(0)f_3^{(3)}\circ\Phi_3^{\tp}(0)\right\rangle^{\tp}_{\mathrm{UHP}}, \label{eq:EB}
\end{equation}
where the conformal transformations 
\[ f^{(3)}_k(z)=h^{-1}\left(e^{\frac{2\pi i}{3}(k-1)}h(z)^{\frac{2}{3}}\right), \quad h(z)=
\frac{1+iz}{1-iz} \]
are generated by the Virasoro operators of the twisted theory. 
The supersliver state $\Xi^{\tp}$ in this twisted conformal field theory is defined by 
\begin{equation}
\langle\Xi^{\tp}|\varphi\rangle=\cN^{\tp}\langle f\circ\varphi^{\tp}(0)\rangle^{\tp}_{\mathrm{UHP}}, 
\label{eq:GA} \\
\end{equation}
with $f(z)=\tan^{-1}z$ for an arbitrary Fock space state $|\varphi\rangle$. $\cN^{\tp}$ is a normalization 
constant which would be needed to achieve $\Xi^{\tp}*^{\tp}\Xi^{\tp}=\Xi^{\tp}$ because of the 
non-vanishing total central charge. The original supersliver $\Xi$ and the twisted supersliver 
$\Xi^{\tp}$ can be written as~\cite{RZ,Wedge}
\begin{eqnarray}
|\Xi\rangle=U_f^{\dagger}|0\rangle, & & U_f^{\dagger}=\exp\left(-\frac{1}{3}L_{-2}+\frac{1}{30}
L_{-4}-\frac{11}{1890}L_{-6}+\cdots\right), \label{eq:GD} \\
|\Xi^{\tp}\rangle=\cN^{\tp}U_f^{\tp\dagger}|0^{\tp}\rangle, & & U_f^{\tp\dagger}=\exp\left(
-\frac{1}{3}L^{\tp}_{-2}+\frac{1}{30}L^{\tp}_{-4}-\frac{11}{1890}L^{\tp}_{-6}+\cdots\right), \label{eq:GE}
\end{eqnarray}
where $L_n (L_n^{\tp})$ are total (\textit{i.e.} matter+ghost) Virasoro generators associated with 
the energy-momentum tensor $T (T^{\tp})$ in the original (twisted) conformal field theory. 
From the relation~(\ref{eq:DG}) between the two $SL(2,\aaru)$ vacua $|0\rangle$ and $|0^{\tp}\rangle$, 
we can rewrite $\Xi^{\tp}$ as 
\begin{equation}
|\Xi^{\tp}\rangle=\cN^{\tp}U_f^{\tp\dagger}\xi ce^{-\phi}(0)|0\rangle =\cN^{\tp}
\left( U_f^{\tp\dagger}\xi ce^{-\phi}(0)U_f^{\dagger -1}\right)|\Xi\rangle . \label{eq:GF}
\end{equation}
From this relation, we find that $\Xi^{\tp}$ has ghost number 0 and picture number 0, and 
is Grassmann odd. (Remember that $\xi$ has ghost number $-1$ and picture number $+1$, and that 
$e^{\ell\phi}$ is Grassmann odd if $\ell$ is an odd integer.) Thus we can consider $\Xi^{\tp}$ 
itself as the string field in the GSO($-$) sector. 
$\Xi^{\tp}*\Xi^{\tp}$ can be calculated by using the formula 
(derived in Appendix~\ref{sec:appA}, eq.(\ref{eq:DT})) 
\begin{equation}
|\Phi_1*\Phi_2\rangle=\lim_{\epsilon\to 0}K^{(3)}
\Big|f^{(3)\prime}_1(i+\epsilon)\Big|^{-\frac{1}{4}}e^{\frac{1}{2}
(\rho-\phi+\chi)(i-\epsilon)}e^{\frac{1}{2}
(\bar{\rho}-\bar{\phi}+\bar{\chi})(i-\epsilon)}|\Phi_1*^{\tp}\Phi_2\rangle \label{eq:GG}
\end{equation}
relating the $*$-product to the $*^{\tp}$-product. Here, $K^{(3)}$ is a normalization constant. 
Since $\Xi^{\tp}$ is a projector of the $*^{\tp}$-algebra, we have 
\begin{equation}
|\Xi^{\tp}*\Xi^{\tp}\rangle=\lim_{\epsilon\to 0}K^{(3)}
\Big|f^{(3)\prime}_1(i+\epsilon)\Big|^{-\frac{1}{4}}
e^{\frac{1}{2}(\rho-\phi+\chi)(i-\epsilon)}e^{\frac{1}{2}(\bar{\rho}-\bar{\phi}+\bar{\chi})
(i-\epsilon)}|\Xi^{\tp}\rangle. \label{eq:GH}
\end{equation}
\smallskip

For later purpose, we collect here some formulas regarding the conformal field theory CFT$^{\prime}$ 
twisted only in the $bc$-ghost sector. Using the correlation function normalized as 
\[ \langle\xi c^{\prime}e^{-2\phi}\rangle^{\prime}_{\mathrm{UHP}}=V_{10}, \]
the $bc$-twisted sliver $\Xi^{\prime}$ is defined by 
\begin{equation}
\langle\Xi^{\prime}|\varphi\rangle=\cN^{\prime}\langle f\circ\varphi^{\prime}(0)
\rangle^{\prime}_{\mathrm{UHP}}, \label{eq:GQ}
\end{equation}
where $\varphi^{\prime}(0)$ is the vertex operator representation of the state $|\varphi\rangle$ 
with respect to the $SL(2,\aaru)$-invariant vacuum $|0^{\prime}\rangle=c(0)|0\rangle$ of the 
$bc$-twisted theory. The $*^{\prime}$-product of this theory is defined as 
\[ \langle\Phi_1|\Phi_2*^{\prime}\Phi_3\rangle=\left\langle f^{(3)}_1\circ\Phi_1^{\prime}(0)
f^{(3)}_2\circ\Phi_2^{\prime}(0)f^{(3)}_3\circ\Phi_3^{\prime}(0)\right\rangle^{\prime}_{\mathrm{UHP}}, \]
and is related to the $*$-product of the original theory through~\cite{GRSZ1} 
\begin{equation}
|\Phi_1*\Phi_2\rangle=\lim_{\epsilon\to 0}K^{(3)}|f^{(3)\prime}_1(i+\epsilon)|^{\frac{1}{4}}
e^{\frac{1}{2}\rho(i-\epsilon)}e^{\frac{1}{2}\bar{\rho}(i-\epsilon)}|\Phi_1*^{\prime}\Phi_2\rangle, 
\label{eq:GW}
\end{equation}

\sectiono{Search for the D-brane Solution: Guesswork}\label{sec:4}
Since we do not know what form the ghost part of the brane solution should take, we will proceed 
by making use of an analogy with bosonic vacuum string field theory. 
By definition, $\widehat{A}(\widehat{\Phi})=e^{-\widehat{\Phi}}\widehat{\cQ}e^{\widehat{\Phi}}$ 
satisfies
\begin{equation}
\widehat{\cQ}\widehat{A}+\widehat{A}*\widehat{A}=0, \label{eq:ER}
\end{equation}
if $\widehat{\cQ}$ is a nilpotent derivation and annihilates the identity $\cI$. Noting that both $\widehat{A}$ 
and $\widehat{\cQ}$ carry ghost number 1 and picture number 0, this `constraint' equation closely resembles 
the equation of motion of bosonic VSFT. Since we already know that the $bc$-twisted sliver $\Xi^{\prime}$ 
(defined below), if normalized suitably, satisfies~\cite{GRSZ1}
\begin{equation}
\Qmid \Xi^{\prime}+\Xi^{\prime}*\Xi^{\prime}=0, \label{eq:ES}
\end{equation}
and is supposed to represent a spacetime-filling D25-brane in bosonic string theory, 
we guess that the spacetime-filling D9-brane and 
the ghost kinetic operator in the superstring case may be given by 
\begin{eqnarray}
\widehat{A}&=&e^{-\widehat{\Phi}}\widehat{\cQ}e^{\widehat{\Phi}}=C
\Xi^{\prime}\otimes\sigma_3, \label{eq:GB} \\
\widehat{\cQ}&=&\frac{1}{2i}(c(i)-c(-i))\otimes\sigma_3=\frac{1}{2}(c^{\prime}(i)+c^{\prime}(-i))\otimes
\sigma_3, \label{eq:ET}
\end{eqnarray}
which have the correct ghost and picture numbers. Here we have tensored the intenal Chan-Paton 
matrices~\cite{TPNS,BSZ}, and the normalization constant $C$ should be uniquely determined by 
the non-linear equation~(\ref{eq:ER}).
The equation of motion $\widehat{\eta}_0\widehat{A}=0$ is solved by the 
configuration~(\ref{eq:GB}) because the $bc$-twisting does not affect the $\xi\eta$-sector at all. 
There is one more reason to identify the specific combination $\widehat{A}$ of the string field 
$\widehat{\Phi}$ in Berkovits' theory with the string field $\Psi$ in cubic string field theory. 
In cubic theory, the new kinetic operator $\cQ^{\prime}$ arising from the expansion around a 
classical solution $\Psi_0$ takes the form 
\begin{equation}
\cQ^{\prime}X=\cQ X+\Psi_0*X-(-1)^{|X|}X*\Psi_0. \label{eq:RC}
\end{equation}
On the other hand, it was shown in~\cite{Klu5,MS} 
that in Berkovits' theory, when the string field $\Phi$ is expanded around a classical 
solution $\Phi_0$ as $e^{\Phi}=e^{\Phi_0}*e^{\phi}$ with $\phi$ denoting the fluctuation field, 
the resultant action for $\phi$ takes the same form as the original one, but with a new 
kinetic operator $\cQ^{\prime}$ given by 
\begin{equation}
\cQ^{\prime}X=\cQ X+A(\Phi_0)*X-(-1)^{|X|}X*A(\Phi_0), \label{eq:CF}
\end{equation}
where $X$ is an arbitrary string field, and $|X|$ is its Grassmannality. 
Eq.(\ref{eq:CF}) looks the same as eq.(\ref{eq:RC}) if we replace $A(\Phi_0)$ with $\Psi_0$. 
Taking into account the facts that the open string spectrum around the classical 
solution is governed by $\cQ^{\prime}$ and that 
seemingly the correct open string spectrum is reproduced around the matter sliver 
solution $(\Psi_0)_{\mathrm{m}}=\Xi_{\mathrm{m}}$ in bosonic VSFT~\cite{HK,HM,RSZ6,HMT,RV,Okawa}, 
we are inclined to take at least the matter part of $A$ to be the matter sliver $\Xi_{\mathrm{m}}$ 
as in (\ref{eq:GB}). This also motivates us to adopt the ansatz~(\ref{eq:GB}),(\ref{eq:ET}). 
\smallskip

The problem now is how to solve the equation~(\ref{eq:GB}) for $\widehat{\Phi}$ in the ``large" Hilbert space. 
Though we have not succeeded in finding an appropriate solution, we will give an example of calculations 
of $e^{-\widehat{\Phi}}\widehat{\cQ}e^{\widehat{\Phi}}$ for a certain string field configuration 
$\widehat{\Phi}_0$ which has some desired properties. Since we are looking for a configuration 
representing a D-brane, we require $\widehat{\Phi}_0$ to satisfy the following conditions: 
\begin{itemize}
\item[(1)] $\widehat{\Phi}_0$ has ghost number 0 and picture number 0, in order for $\widehat{\Phi}_0$ 
to be acceptable as a string field;
\item[(2)] $\widehat{\Phi}_0$ has a matter-ghost split form and its matter part is simply the 
matter sliver $\Xi_{\mathrm{m}}$;
\item[(3)] $e^{-\widehat{\Phi}_0}\widehat{\cQ}e^{\widehat{\Phi}_0}$ is exactly calculable;
\item[(4)] $\widehat{\Phi}_0$ is not a pure-gauge configuration. In particular, $\widehat{\Phi}_0$ 
should not be killed by $\widehat{\eta}_0$ nor $\widehat{\cQ}$.
\end{itemize}
It turns out that the triply-twisted supersliver state $\Xi^{\tp}$ introduced in the previous section 
satisfies the above conditions. So we try  
\begin{equation}
\widehat{\Phi}_0=a_-\Xi^{\tp}\otimes\sigma_1, \label{eq:GC}
\end{equation}
where we have attached the internal Chan-Paton matrix $\sigma_1$ according to its GSO parity (\textit{i.e.} 
Grassmannality). The conditions (1),(2) are satisfied, as noticed in section~\ref{sec:3}. 
The condition (4) can be shown in the following way: 
We can show that $\Xi^{\tp}$ is now annihilated 
by $\xi_0^{\prime}$ in the same way as in eq.(\ref{eq:CG}), because $\xi^{\prime}(z)$ is a primary field 
of comformal weight 1 in the twisted theory. Given that $\{\xi^{\prime}_0,\eta^{\prime}_0\}=1$, it follows that 
$\Xi^{\tp}$ must not be killed by $\eta^{\prime}_0=\eta_0$. On the other hand, 
\[ \langle\Xi^{\tp}|\cQ|\varphi\rangle=\left\langle\frac{1}{2}f\circ(c^{\prime}(i)+c^{\prime}(-i))
f\circ\varphi^{\tp}(0)\right\rangle^{\tp}_{\mathrm{UHP}}=\langle c^{\prime}(\infty)\ f\circ\varphi^{\tp}(0)
\rangle^{\tp}_{\mathrm{UHP}} \]
does not vanish for a general state $|\varphi\rangle$ 
because $c^{\prime}(\pm i)$ in the twisted theory has 
conformal weight 0 so that they give rise to no vanishing conformal factors. 
\smallskip

Now we calculate $\widehat{A}(\widehat{\Phi}_0)=e^{-\widehat{\Phi}_0}\widehat{\cQ}e^{\widehat{\Phi}_0}$. 
Making use of (\ref{eq:GH}) $(n-1)$ times and (\ref{eq:DW}), we obtain 
\begin{equation}
\overbrace{\Xi^{\tp}*\ldots *\Xi^{\tp}}^n=(K^{(3)})^{n-1}
|f^{(3)\prime}_1(i)|^{-\frac{n-1}{4}}\left(e^{\frac{1}{2}
(\rho-\phi+\chi)(i)}e^{\frac{1}{2}(\bar{\rho}-\bar{\phi}+\bar{\chi})(i)}\right)^{n-1}|\Xi^{\tp}\rangle, 
\label{eq:M92C}
\end{equation}
where $\epsilon$-regularization is implicit, and the $(n-1)$-fold operator product above is \textit{not} 
normal ordered yet. We have assumed that the associativity of the $*$-product is not 
violated by $\Xi^{\tp}$. We can write $e^{\widehat{\Phi}_0}$ in the form 
\begin{eqnarray}
e^{\widehat{\Phi}_0}&=&\cI\otimes\mathbf{1}+\sum_{n=1}^{\infty}\frac{1}{n!}\overbrace{\widehat{\Phi}_0
*\ldots *\widehat{\Phi}_0}^n \nonumber \\
&=&\cI\otimes\mathbf{1}+\sum_{n=1}^{\infty}\frac{a_-^n}{n!}(K^{(3)})^{n-1}
|f^{(3)\prime}_1(i)|^{-\frac{n-1}{4}} 
\label{eq:GX} \\ & &{}\quad\times\left(e^{\frac{1}{2}(\rho-\phi+\chi)(i)}e^{\frac{1}{2}(\bar{\rho}
-\bar{\phi}+\bar{\chi})(i)}\right)^{n-1}
|\Xi^{\tp}\rangle\otimes\sigma_1^n. \nonumber 
\end{eqnarray}
If the kinetic operator~(\ref{eq:ET}) is regularized as
\begin{equation}
\widehat{\cQ}=\lim_{\epsilon\to 0}\frac{1}{2}\left(\cQ^{\epsilon}-I\circ \cQ^{\epsilon}\right)\otimes
\sigma_3 \qquad 
\mathrm{with} \quad \cQ^{\epsilon}=\frac{1}{2i}\left(e^{-i\epsilon}c(ie^{i\epsilon})-e^{i\epsilon}
c(-ie^{-i\epsilon})\right), \label{eq:CE}
\end{equation}
where $I(z)=-1/z$, then $\widehat{\cQ}$ annihilates 
the identity $\cI$ even for finite $\epsilon$~\cite{GRSZ1,KiOh}. 
On the sliver state twisted in the $bc$-sector, $\Qmid$ can be rewritten as~\cite{GRSZ1} 
\begin{equation}
\Qmid|\Xi^{\prime}\rangle=|(h\circ f)^{\prime}(i)|^{1/4}|h\circ f(i)-h\circ f(-i)|^{-1/4}
e^{\frac{1}{2}\rho(i)}e^{\frac{1}{2}\rho(-i)}|\Xi^{\prime}\rangle, \label{eq:RH}
\end{equation}
because 
\begin{eqnarray*}
& &\hspace{-1.5cm}\langle\Xi^{\prime}|e^{\frac{1}{2}\rho(i)}e^{\frac{1}{2}\rho(-i)}|\varphi\rangle=
\langle f\circ (e^{\frac{1}{2}\rho(i)}e^{\frac{1}{2}\rho(-i)})f\circ\varphi^{\prime}(0)
\rangle^{\prime}_{\mathrm{UHP}} \\ &=&\langle h\circ f\circ(e^{\frac{1}{2}\rho(i)}e^{\frac{1}{2}\rho(-i)})\ 
h\circ f\circ \varphi^{\prime}(0)\rangle^{\prime}_{\mathrm{disk}} \\
&=&|(h\circ f)^{\prime}(i)|^{-\frac{1}{4}}|h\circ f(i)-h\circ f(-i)|^{\frac{1}{4}}\langle e^{\rho
(h\circ f(i))}\ h\circ f\circ\varphi^{\prime}(0)\rangle^{\prime}_{\mathrm{disk}} \\
&=&|(h\circ f)^{\prime}(i)|^{-\frac{1}{4}}|h\circ f(i)-h\circ f(-i)|^{\frac{1}{4}}\langle c^{\prime}(\infty)
\ f\circ\varphi^{\prime}(0)\rangle^{\prime}_{\mathrm{UHP}} \\
&=&|(h\circ f)^{\prime}(i)|^{-\frac{1}{4}}|h\circ f(i)-h\circ f(-i)|^{\frac{1}{4}}\langle\Xi^{\prime}|
\Qmid|\varphi\rangle, 
\end{eqnarray*}
where we have used the conformal map $h(z)=\frac{1+iz}{1-iz}$ to rewrite the correlator on the upper half plane 
as the disk correlator. 
Eq.(\ref{eq:RH}) holds for $|\Xi^{\tp}\rangle$ as well. From these facts, we find 
\begin{eqnarray}
\widehat{\cQ}e^{\widehat{\Phi}_0}&=&
|(h\circ f)^{\prime}(i)|^{\frac{1}{4}}|h\circ f(i)-h\circ f(-i)|^{-\frac{1}{4}}
\sum_{n=1}^{\infty}\frac{a_-^n}{n!} (K^{(3)})^{n-1} |
f^{(3)\prime}_1(i)|^{-\frac{n-1}{4}}\nonumber \\ & &{}\quad\times\left( e^{\frac{1}{2}\rho(i)}
e^{\frac{1}{2}\bar{\rho}(i)}\right)^n  
\left(e^{\frac{1}{2}(-\phi+\chi)(i)}e^{\frac{1}{2}(
-\bar{\phi}+\bar{\chi})(i)}\right)^{n-1}|\Xi^{\tp}\rangle\otimes\sigma_3\sigma_1^n. \label{eq:GY}
\end{eqnarray}
$e^{-\widehat{\Phi}_0}$ can easily be found by replacing $a_{-}$ with $-a_{-}$ in (\ref{eq:GX}). Then we have 
\begin{eqnarray}
& &\hspace{-1cm}\widehat{A}(\widehat{\Phi}_0)=e^{-\widehat{\Phi}_0}\widehat{\cQ}e^{\widehat{\Phi}_0} \nonumber \\
&=&|(h\circ f)^{\prime}(i)|^{\frac{1}{4}}|h\circ f(i)-h\circ f(-i)|^{-\frac{1}{4}}
\sum_{n=1}^{\infty}\sum_{n^{\prime}=0}^{\infty}\frac{a_-^{n+n^{\prime}}}{n!n^{\prime}!}
(K^{(3)})^{n+n^{\prime}-1} \label{eq:HA} \\
& &\ \ {}\times |f^{(3)\prime}_1(i)|^{-\frac{n+n^{\prime}-1}{4}} 
\left( e^{\frac{1}{2}\rho(i)}e^{\frac{1}{2}\bar{\rho}(i)}\right)^{n+n^{\prime}}
\left(e^{\frac{1}{2}(-\phi+\chi)(i)}e^{\frac{1}{2}(-\bar{\phi}+\bar{\chi})(i)}
\right)^{n+n^{\prime}-1}|\Xi^{\tp}\rangle\otimes\sigma_3\sigma_1^{n+n^{\prime}}. 
\nonumber
\end{eqnarray}
Thus we have exactly calculated $e^{-\widehat{\Phi}_0}\widehat{\cQ}e^{\widehat{\Phi}_0}$ 
for $\widehat{\Phi}_0$ given in~(\ref{eq:GC}), up to the cocycle factors 
for the exponential operators we have been ignoring. 
Since this configuration turns out not to solve 
the equation of motion $\widehat{\eta}_0\widehat{A}=0$ unless $a_-=0$, 
we cannot consider the triply-twisted supersliver $\Xi^{\tp}$ itself as 
representing a D-brane. However, considering the properties possessed by $\Xi^{\tp}$, 
we expect that it may be relevant 
to the construction of the correct D-brane solutions.
\smallskip

As the second example, we consider the string field configuration 
\begin{equation}
\widehat{\Phi}_e=a_+e^{-\frac{1}{2}\rho(i)}e^{-\frac{1}{2}\bar{\rho}(i)}|\Xi^{\prime}\rangle
\otimes\mathbf{1}, \label{eq:M92A}
\end{equation}
though it is annihilated by $\eta_0$ and hence it is pure-gauge. 
$\widehat{\Phi}_e$ is Grassmann-even just 
like the untwisted sliver, and has ghost number 0 and picture number 0, as it should be.  
From eqs.(\ref{eq:DW}) and (\ref{eq:GW}), we find 
\begin{equation}
\widehat{\Phi}_e *\widehat{\Phi}_e=\lim_{\epsilon\to 0}a_+K^{(3)}
|f^{(3)\prime}_1(i+\epsilon)|^{\frac{1}{4}}
\left(e^{-\frac{1}{2}\rho(i)}e^{-\frac{1}{2}\bar{\rho}(i)}\cdot 
e^{\frac{1}{2}\rho(i-\epsilon)}
e^{\frac{1}{2}\bar{\rho}(i-\epsilon)}\right)\widehat{\Phi}_e. \label{eq:M92B}
\end{equation}
In the $\epsilon\to 0$ limit, $e^{-\frac{1}{2}\rho(i)}e^{\frac{1}{2}\rho(i-\epsilon)}$ 
can be replaced by the leading term $\epsilon^{-\frac{1}{4}}$ of the OPE. 
Using~(\ref{eq:M92B}) recursively, we obtain 
\begin{eqnarray}
& &\hspace{-1.5cm}\widehat{A}(\widehat{\Phi}_e)=e^{-\Phi_e}\cQ e^{\Phi_e}\otimes \sigma_3 \nonumber \\
&=& \lim_{\epsilon\to 0}
|(h\circ f)^{\prime}(i+\epsilon)|^{\frac{1}{4}}|h\circ f(i+\epsilon)-h\circ f(-i-\epsilon)|^{-\frac{1}{4}}
a_+\epsilon^{-\frac{1}{2}}\nonumber \\ & &\times\sum_{n=1}^{\infty}
\sum_{n^{\prime}=0}^{\infty}\frac{(-1)^{n^{\prime}}}{n!n^{\prime}!}\left(a_+K^{(3)}
|f^{(3)\prime}_1(i+\epsilon)|^{\frac{1}{4}}\epsilon^{-\frac{1}{2}}\right)^{n+n^{\prime}-1}
\Xi^{\prime}\otimes\sigma_3 \nonumber \\
&=&\lim_{\epsilon\to 0}\frac{1}{K^{(3)}}|f^{(3)\prime}_1(i+\epsilon)|^{-\frac{1}{4}}
|(h\circ f)^{\prime}(i+\epsilon)|^{\frac{1}{4}}|h\circ f(i+\epsilon)-h\circ f(-i-\epsilon)|^{-\frac{1}{4}}\nonumber \\ & &\times
\left( 1-\exp\left(-a_+K^{(3)}|f^{(3)\prime}_1(i+\epsilon)|^{\frac{1}{4}}\epsilon^{-\frac{1}{2}}
\right)\right)\Xi^{\prime}\otimes\sigma_3  \label{eq:RF} \\
&\propto&\Xi^{\prime}\otimes\sigma_3. \nonumber
\end{eqnarray}
in the same way as in (\ref{eq:M92C})--(\ref{eq:HA}).
From this expression, one may think that we have obtained the solution 
$\widehat{A}=C\Xi^{\prime}\otimes\sigma_3$~(\ref{eq:GB}) from 
$\widehat{\Phi}_e$, but unfortunately 
this configuration is pure-gauge, as mentioned above. So we have to say that it 
seems difficult to obtain the \textit{non-trivial} solution~(\ref{eq:GB}) in vacuum 
superstring field theory with $\widehat{\cQ}$ given in~(\ref{eq:ET}).

\sectiono{Remarks on Bosonic VSFT Solutions}\label{sec:5}
In this section we turn our attention to the bosonic case. 
In bosonic vacuum string field theory, it was demonstrated in~\cite{GRSZ1} that the equation of motion 
\begin{equation}
\Qmid \Psi + \Psi * \Psi = 0 \label{eq:RG}
\end{equation}
is solved by the $bc$-twisted sliver. Plugging $\Psi=C\Xi^{\prime}$ into (\ref{eq:RG}), 
we find the left hand side of (\ref{eq:RG}) to be 
\begin{equation}
C\left(1+CK^{(3)}|f^{(3)\prime}_1(i)|^{\frac{1}{4}}|(h\circ f)^{\prime}(i)|^{-\frac{1}{4}}
|h\circ f(i)-h\circ f(-i)|^{\frac{1}{4}}\right)\Qmid|\Xi^{\prime}\rangle, \label{eq:RI}
\end{equation}
where we have used (\ref{eq:RH}) and (\ref{eq:GW}), the latter of which is valid also in the 
bosonic case. By requiring~(\ref{eq:RI}) to vanish, the overall normalization of the 
solution is uniquely fixed as 
\begin{equation}
\Psi_0=-\frac{1}{K^{(3)}|f^{(3)\prime}_1(i+\epsilon)|^{1/4}|(h\circ f)^{\prime}(i+\epsilon)|^{-1/4}
|h\circ f(i+\epsilon)-h\circ f(-i-\epsilon)|^{1/4}}\Xi^{\prime}. \label{eq:RJ}
\end{equation}
Recalling the result~(\ref{eq:RF}) from the last section, we have 
\begin{eqnarray}
& &\hspace{-1cm}\exp\left(-a_+e^{-\frac{1}{2}\rho(i)}e^{-\frac{1}{2}\bar{\rho}(i)}\Xi^{\prime}\right)\Qmid 
\exp\left(a_+e^{-\frac{1}{2}\rho(i)}e^{-\frac{1}{2}\bar{\rho}(i)}\Xi^{\prime}\right) \nonumber \\
&=&\lim_{\epsilon\to 0}\frac{1}{K^{(3)}}|f^{(3)\prime}_1(i+\epsilon)|^{-1/4}
|(h\circ f)^{\prime}(i+\epsilon)|^{1/4}|h\circ f(i+\epsilon)-h\circ f(-i-\epsilon)|^{-1/4}\nonumber \\ & &\times
\left( 1-\exp\left(-a_+K^{(3)}|f^{(3)\prime}_1(i+\epsilon)|^{\frac{1}{4}}\epsilon^{-\frac{1}{2}}
\right)\right)\Xi^{\prime}. \label{eq:RN}
\end{eqnarray}
Comparing (\ref{eq:RN}) with (\ref{eq:RJ}), we find that if $a_+$ is determined such that 
\begin{equation}
1-\exp\left(-a_+K^{(3)}|f^{(3)\prime}_1(i+\epsilon)|^{\frac{1}{4}}\epsilon^{-\frac{1}{2}}
\right)=-1 \label{eq:RO}
\end{equation}
is satisfied, then the solution $\Psi_0$ can be written as 
\begin{equation}
\Psi_0=\exp\left(-a_+e^{-\frac{1}{2}\rho(i)}e^{-\frac{1}{2}\bar{\rho}(i)}\Xi^{\prime}\right)\Qmid 
\exp\left(a_+e^{-\frac{1}{2}\rho(i)}e^{-\frac{1}{2}\bar{\rho}(i)}\Xi^{\prime}\right). \label{eq:RP}
\end{equation}

On the other hand, finite gauge transformation in 
cubic string field theory is generally given by 
\begin{equation}
\Psi \quad \longrightarrow \quad e^{-\Lambda}(Q+\Psi)e^{\Lambda}, \label{eq:RL}
\end{equation}
where $Q$ is the kinetic operator of the theory and $\Lambda$ is a gauge parameter of ghost number 0. 
Then it follows that the string field configuration of the form 
\begin{equation}
e^{-\Lambda}Qe^{\Lambda} \label{eq:RM}
\end{equation}
is pure-gauge. 
Since the expression~(\ref{eq:RP}) 
is of the form (\ref{eq:RM}) with $Q=\Qmid$ and 
\begin{equation}
\Lambda=-\frac{\log 2}{K^{(3)}}
\lim_{\epsilon\to 0}\epsilon^{\frac{1}{2}}|f^{(3)\prime}_1(i+\epsilon)|^{-\frac{1}{4}}e^{-\frac{1}{2}
\rho(i+\epsilon)}e^{-\frac{1}{2}\bar{\rho}(i+\epsilon)}\Xi^{\prime}, \label{eq:M92D}
\end{equation}
it seems that the $bc$-twisted sliver solution $\Psi_0$ is pure-gauge in bosonic VSFT with 
$\cQ=\Qmid$. However, this conclusion holds true only when the above $\Lambda$~(\ref{eq:M92D}) 
is allowed as a 
gauge transformation parameter. In fact, it was argued in~\cite{RSZ4} that 
the gauge parameters having logarithmically divergent norms in the matter sector should 
not be allowed around a D-brane solution. In our case, we notice the following fact: 
As we mentioned at the beginning of section~\ref{sec:4}, string field configurations $\psi$ of the 
form~(\ref{eq:RM}) should always satisfy an equation $Q\psi+\psi *\psi=0$ irrespective of 
details of $\Lambda$, as long as $Q$ is a nilpotent derivation. Contrary to this general argument, 
however, the configuration given in eq.(\ref{eq:RP}) does not obey this equation unless $a_+$ 
satisfies the condition~(\ref{eq:RO}). 
This might be a manifestation of the fact that $\Lambda$ in~(\ref{eq:M92D}) is ill-defined as a 
gauge parameter so that it violates some of the formal properties the allowed gauge parameters 
should have. Thus, it is possible that the solution 
$\Psi_0$ is actually \textit{not} a pure-gauge 
configuration in bosonic VSFT, in spite of the expression~(\ref{eq:RP}). 
Nevertheless, we believe that in the superstring case 
considered in section~\ref{sec:2} the string field configurations annihilated by $\eta_0$ are 
really pure-gauge because of the circumstantial evidence that such configurations are tensionless 
and form a continuous family of solutions including $\Phi=0$.

\sectiono{Summary and Discussion}\label{sec:sum}
In this paper we have presented several pieces of evidence that the untwisted supersliver solution 
is not appropriate for the description of a D-brane in non-polynomial vacuum superstring field theory. 
In particular, we have shown that any state 
annihilated by $\eta_0$ (or $\cQ$) is non-perturbatively pure-gauge. 
We have then guessed the possible form of the kinetic operator $\widehat{\cQ}$~(\ref{eq:ET}) and 
the spacetime-filling D-brane solution $\widehat{A}$~(\ref{eq:GB}) in this theory, 
based on an analogy with those of bosonic vacuum string field theory. 
As an effort at constructing non-trivial solutions, 
we have used the triply-twisted supersliver $\Xi^{\tp}$, which is not a gauge degree of freedom, 
to give an example in which $e^{-\widehat{\Phi}} 
\widehat{\cQ}e^{\widehat{\Phi}}$ can be calculated exactly. 
It may turn out that our proposals~(\ref{eq:GB})--(\ref{eq:ET}) are not quite correct, especially 
for the reason discussed below. Nevertheless, we hope that our ideas 
as well as the computational techniques developed in this paper 
will be useful in clarifying the structure of vacuum superstring field theory solutions. 

As a by-product of the above calculations, we have found that the $bc$-twisted sliver solution 
in bosonic vacuum string field theory, which has been 
believed to be the D-brane solution, 
can formally be written as a pure-gauge configuration. 
We have not reached a definite conclusion on this point because we have not determined whether 
the required gauge transformation is a valid one or not. 
We end this paper by making a few remarks on further research. 
\medskip

From the internal Chan-Paton structure 
\[ \widehat{\cQ}=\cQ\otimes\sigma_3,\quad \widehat{\eta}_0=\eta_0\otimes\sigma_3,\quad
\widehat{\Phi}=\Phi_+\otimes\mathbf{1}+\Phi_-\otimes\sigma_1, \]
we can see that the action 
\begin{equation}
S=\frac{\kappa_0}{2}\mathrm{Tr}\int\left(\left(e^{-\widehat{\Phi}}\widehat{\cQ} e^{\widehat{\Phi}}\right)
\left(e^{-\widehat{\Phi}}\widehat{\eta}_0 e^{\widehat{\Phi}}
\right)-\int\limits_0^1dt\left(e^{-t\widehat{\Phi}}\partial_te^{t\widehat{\Phi}}\right)\left\{
e^{-t\widehat{\Phi}}\widehat{\cQ} e^{t\widehat{\Phi}},
e^{-t\widehat{\Phi}}\widehat{\eta}_0e^{t\widehat{\Phi}}\right\}\right),
\end{equation}
is invariant under the $\zetto_2$-transformation 
\begin{equation}
\Phi_+\otimes\mathbf{1}+\Phi_-\otimes\sigma_1\quad\longrightarrow\quad
\Phi_+\otimes\mathbf{1}-\Phi_-\otimes\sigma_1.
\label{eq:IA}
\end{equation}
It then follows that, if we have a solution with a non-vanishing GSO($-$) component, there always exists 
one more solution with the same energy density. Since we do not anticipate such a two-fold degeneracy 
of solutions in vacuum superstring field theory~\cite{RSZ1}, this result may imply that we should modify 
the kinetic operator as 
\begin{equation}
\widehat{\cQ}=\cQ_{\mathrm{odd}}\otimes\sigma_3+\cQ_{\mathrm{even}}\otimes i\sigma_2, \label{eq:IB}
\end{equation}
as proposed in~\cite{ABG}. Here $\cQ_{\mathrm{odd}}$ and $\cQ_{\mathrm{even}}$ are Grassmann odd and 
even operators, respectively. It would be interesting to further investigate this possibility. 
\smallskip

Aside from other D-brane solutions which appear in type IIA superstring theory, we should 
be able to construct `another vacuum' solution with vanishing energy density, which corresponds 
to the other minimum of the double-well potential. 
This solution may be given by a configuration $\widehat{\Phi}$ with $\widehat{A}(\widehat{\Phi})=0$ 
because we expect the new kinetic operator $\cQ^{\prime}$~(\ref{eq:CF}) around this vacuum to have 
zero cohomology as well. Since the solutions representing this vacuum or BPS D-branes (brane-antibrane 
pairs) have no analogues in bosonic string theory, to seek them would be a challenging problem. 

\section*{Acknowledgements}
I am grateful to I. Kishimoto for careful reading of the manuscript and many helpful discussions. 
I'd also like to thank T. Eguchi, T. Kawano, K. Sakai, T. Suyama, T. Takayanagi and S. Teraguchi   
for useful discussions and stimulating conversations. This work is supported by JSPS Research 
Fellowships for Young Scientists. 


\section*{Appendices}
\renewcommand{\thesection}{\Alph{section}}
\setcounter{section}{0}

\sectiono{Relating $*$ and $*^{\tp}$}\label{sec:appA}
In this appendix, we derive the formula~(\ref{eq:GG}) relating the $*$-product and $*^{\tp}$-product 
with some details, closely following the strategy of~\cite{GRSZ1}. The $n$-string interaction vertex of 
string field theory is defined as~\cite{Witten,LPP}
\begin{equation}
\int\Phi_1*\ldots *\Phi_n=\langle\Phi_1|\Phi_2*\ldots *\Phi_n\rangle=\langle f^{(n)}_1\circ\Phi_1(0)
\ldots f^{(n)}_n\circ\Phi_n(0)\rangle_{\mathrm{UHP}}, \label{eq:DJ}
\end{equation}
using the correlation function in the \textit{original} CFT. Here, the conformal maps are 
given by 
\begin{equation}
f^{(n)}_k(z)=h^{-1}\left(e^{\frac{2\pi i}{n}(k-1)}h(z)^{\frac{2}{n}}\right), \quad  
h(z)=\frac{1+iz}{1-iz}. \label{eq:DK}
\end{equation}
Now let us consider rewriting the expression~(\ref{eq:DJ}) in terms of the twisted CFT$^{\tp}$. 
When the $n$ strips are glued together with the Witten's prescription~\cite{Witten}, the resultant 
surface is flat everywhere except at the common midpoint of the $n$ open strings, where we have an 
excess angle $(n-2)\pi$. Substituting $R=-(n-2)\pi\delta^{(2)}(w-\frac{\pi}{2})$ into eq.(\ref{eq:DD}), 
we find that $\cS$ and $\cS^{\tp}$ are related in this case as 
\begin{equation}
\cS^{\tp}=\cS+\frac{n-2}{2}(\rho+\bar{\rho}-\phi-\bar{\phi}+\chi+\bar{\chi})(z=-e^{-iw}=i). \label{eq:Mem}
\end{equation}
In the path integral formulation, the correlator~(\ref{eq:DJ}) 
is written as 
\begin{equation}
\langle f^{(n)}_1\circ\Phi_1(0)\ldots f^{(n)}_n\circ\Phi_n(0)\rangle=\frac{V_{10}}{Z}\int\cD\omega
f^{(n)}_1\circ\Phi_1(0)\ldots f^{(n)}_n\circ\Phi_n(0)e^{-\cS[\omega]}, \label{eq:DN}
\end{equation}
where $\omega$ collectively denotes the world-sheet fields in the original theory, \\
$Z=\int\cD\omega\xi\frac{1}{2}c\partial c\partial^2ce^{-2\phi}e^{-\cS}$ is the normalization 
factor. Using the relations (\ref{eq:Mem}) 
and (\ref{eq:DH}), the above expression~(\ref{eq:DN}) can be rewritten as 
\begin{eqnarray}
& &\frac{V_{10}}{Z}\int\cD\omega f^{(n)}_1\circ\Phi_1(0)\ldots f^{(n)}_n\circ\Phi_n(0)e^{-\cS[\omega]}
\nonumber \\ &\simeq& \frac{V_{10}}{Z^{\tp}}\int\cD\omega^{\tp}f^{(n)}_1\circ\Phi_1^{\tp}(0)
\ldots f^{(n)}_n\circ\Phi_n^{\tp}(0)e^{\frac{n-2}{2}(\rho-\phi+\chi)(i)}
e^{\frac{n-2}{2}(\bar{\rho}-\bar{\phi}+\bar{\chi})(i)}
e^{-\cS^{\tp}[\omega^{\tp}]} \nonumber \\
&=&\left\langle f^{(n)}_1\circ\Phi_1^{\tp}(0)\ldots f^{(n)}_n\circ\Phi_n^{\tp}(0)
e^{\frac{n-2}{2}(\rho-\phi+\chi)(i)}e^{\frac{n-2}{2}
(\bar{\rho}-\bar{\phi}+\bar{\chi})(i)}\right\rangle^{\tp}. \label{eq:DO}
\end{eqnarray}
Thus we have obtained 
\begin{eqnarray}
& &\langle f^{(n)}_1\circ\Phi_1(0)\ldots f^{(n)}_n\circ\Phi_n(0)\rangle_{\mathrm{UHP}} \nonumber \\
&=&K^{(n)}
\left\langle f^{(n)}_1\circ\Phi_1^{\tp}(0)\ldots f^{(n)}_n\circ\Phi_n^{\tp}(0)
e^{\frac{n-2}{2}(\rho-\phi+\chi)(i)}e^{\frac{n-2}{2}
(\bar{\rho}-\bar{\phi}+\bar{\chi})(i)}\right\rangle^{\tp}_{\mathrm{UHP}}, \label{eq:DP}
\end{eqnarray}
where we have put a finite normalization constant $K^{(n)}$. 
Since $f^{(n)}_1(i)=i$ and 
$e^{\frac{n-2}{2}(\rho-\phi+\chi)}e^{\frac{n-2}{2}(\bar{\rho}-\bar{\phi}+\bar{\chi})}$ 
is a primary field of conformal weight $(d_n,d_n)=((n-2)^2/8,(n-2)^2/8)$ in the 
twisted theory, we can further rewrite (\ref{eq:DP}) as (up to a possible phase) 
\begin{equation}
K^{(n)}|f^{(n)\prime}_1(i)|^{-2d_n}\left\langle f_1^{(n)}\circ\left(\Phi^{\tp}_1(0)
e^{\frac{n-2}{2}(\rho-\phi+\chi)(i)}e^{\frac{n-2}{2}(
\bar{\rho}-\bar{\phi}+\bar{\chi})(i)}\right) f_2^{(n)}\circ\Phi_2^{\tp}(0)\ldots f_n^{(n)}\circ
\Phi^{\tp}_n(0)\right\rangle_{\mathrm{UHP}}^{\tp}. \label{eq:DQ}
\end{equation}
Here we introduce the $*^{\tp}$-product by the following relation 
\begin{equation}
\langle\Phi_1|\Phi_2*^{\tp}\ldots*^{\tp}\Phi_n\rangle\equiv\left\langle f_1^{(n)}\circ\Phi_1^{\tp}
(0)\ldots f_n^{(n)}\circ\Phi^{\tp}_n(0)\right\rangle^{\tp}_{\mathrm{UHP}}, \label{eq:DR}
\end{equation}
using the \textit{twisted} CFT$^{\tp}$. Comparing (\ref{eq:DQ}) with
\[ \langle\Phi_1|\cO^{\tp}|\Phi_2*^{\tp}\ldots *^{\tp}\Phi_n\rangle=\left\langle
f_1^{(n)}\circ\left(\Phi_1^{\tp}(0)I\circ\cO^{\tp}\right) f_2^{(n)}\circ\Phi_2^{\tp}(0)
\ldots f_n^{(n)}\circ\Phi_n^{\tp}(0)\right\rangle^{\tp}_{\mathrm{UHP}}, \]
where $\cO^{\tp}$ denotes some operator and $I(z)=-1/z$, we find the following formula relating 
the $*$-product~(\ref{eq:DJ}) and the $*^{\tp}$-product~(\ref{eq:DR}), 
\begin{equation}
\langle\Phi_1|\Phi_2*\ldots *\Phi_n\rangle=K^{(n)}|f_1^{(n)\prime}(i)|^{-\frac{(n-2)^2}{4}}\langle\Phi_1|
I\circ \left(e^{\frac{n-2}{2}(\rho-\phi+\chi)(i)}e^{\frac{n-2}{2}(\bar{\rho}-\bar{\phi}+\bar{\chi})(i)}
\right)|\Phi_2*^{\tp}\ldots *^{\tp}
\Phi_n\rangle \label{eq:DS}
\end{equation}
via eqs.(\ref{eq:DP}), (\ref{eq:DQ}). Removing the same state $\langle\Phi_1|$ from both sides, 
we at last reach the desired expression 
\begin{equation}
|\Phi_2*\ldots *\Phi_n\rangle=\lim_{\epsilon\to 0}K^{(n)}|f_1^{(n)\prime}(i+\epsilon)|^{-\frac{(n-2)^2}{4}}
e^{\frac{n-2}{2}(\rho-\phi+\chi)(i-\epsilon)}e^{\frac{n-2}{2}(\bar{\rho}-\bar{\phi}+\bar{\chi})
(i-\epsilon)}|\Phi_2*^{\tp}\ldots *^{\tp}
\Phi_n\rangle \label{eq:DT}
\end{equation}
where we have regularized the expression by introducing a regulation parameter $\epsilon$ because 
\[ f^{(n)\prime}_1(i+\epsilon)\simeq -\frac{4i}{n}\left(\frac{i}{2}\right)^{\frac{2}{n}}
\epsilon^{\frac{2}{n}-1} \]
is divergent in the limit $\epsilon\to 0$ when $n\ge 3$.

\sectiono{Operator Insertion at the Midpoint}\label{sec:appB}
We consider the following expression 
\begin{equation}
\cO(i)\overline{\cO}(i)|\Phi_2*\ldots *\Phi_n\rangle, \label{eq:DU}
\end{equation}
where $\cO(i)$ is assumed to be a holomorphic primary field of conformal weight $(d,0)$ and 
$\overline{\cO}(i)$ is the corresponding antiholomorphic field of weight $(0,d)$. Open string 
boundary condition requires $\overline{\cO}(i)=\cO(-i)$. Taking the inner product of~(\ref{eq:DU}) 
with $\langle\Phi_1|$, we have 
\begin{eqnarray}
& &\langle\Phi_1|\cO(i)\overline{\cO}(i)|\Phi_2*\ldots *\Phi_n\rangle \nonumber \\
&=&\left\langle f_1^{(n)}\circ
\left(\Phi_1(0)I\circ\cO(i)I\circ\overline{\cO}(i)\right)f_2^{(n)}\circ\Phi_2(0)\ldots
f_n^{(n)}\circ\Phi_n(0)\right\rangle_{\mathrm{UHP}} \nonumber \\
&=&|I^{\prime}(i)|^{2d}|f^{(n)\prime}_1(i)|^{2d}\left\langle f_1^{(n)}\circ\Phi_1(0)
\cO(i)\overline{\cO}(i)f_2^{(n)}\circ\Phi_2(0)\ldots f_n^{(n)}\circ\Phi_n(0)\right\rangle_{\mathrm{UHP}}.
\label{eq:DV}
\end{eqnarray}
Since $|I^{\prime}(i)|=1, f^{(n)}_k(i)=i$ for all $k$ and $|f_1^{(n)\prime}(i)|=\ldots =|f_n^{(n)\prime}
(i)|$, we find 
\begin{eqnarray*}
& &\left\langle f_1^{(n)}\circ\left(\Phi_1(0)I\circ\cO(i)I\circ\overline{\cO}(i)\right)f_2^{(n)}
\circ\Phi_2(0)f_3^{(n)}\circ\Phi_3(0)\ldots f_n^{(n)}\circ\Phi_n(0)\right\rangle_{\mathrm{UHP}} \\
&=&\left\langle f_1^{(n)}\circ\Phi_1(0)f_2^{(n)}\circ\left(\cO(i)\overline{\cO}(i)\Phi_2(0)\right)
f_3^{(n)}\circ\Phi_3(0)\ldots f_n^{(n)}\circ\Phi_n(0)\right\rangle_{\mathrm{UHP}} \\
&=&(-1)^{|\cO\overline{\cO}||\Phi_2|}\left\langle f_1^{(n)}\circ\Phi_1(0)f_2^{(n)}\circ\Phi_2(0)
f_3^{(n)}\circ\left(\cO(i)\overline{\cO}(i)\Phi_3(0)\right)\ldots f_n^{(n)}\circ\Phi_n(0)
\right\rangle_{\mathrm{UHP}} \\
&=&\ldots \\
&=&(-1)^{|\cO\overline{\cO}|(|\Phi_2|+\ldots+|\Phi_{n-1}|)}\left\langle f_1^{(n)}\circ\Phi_1(0)
\ldots f_n^{(n)}\circ\left(\cO(i)\overline{\cO}(i)\Phi_n(0)\right)\right\rangle_{\mathrm{UHP}}, 
\end{eqnarray*}
which implies 
\begin{eqnarray}
& &\cO(i)\overline{\cO}(i)|\Phi_2*\ldots *\Phi_n\rangle=\Big|\left(\cO(i)\overline{\cO}(i)\Phi_2\right)
*\Phi_3*\ldots *\Phi_n\Bigr\rangle \nonumber \\
&=&(-1)^{|\cO\overline{\cO}||\Phi_2|}\Big|\Phi_2*\left(\cO(i)\overline{\cO}(i)\Phi_3\right)*\ldots *
\Phi_n\Bigr\rangle=\ldots \label{eq:DW} \\
&=&(-1)^{|\cO\overline{\cO}|(|\Phi_2|+\ldots+|\Phi_{n-1}|)}\Big|\Phi_2*\ldots *\Phi_{n-1}*
\left(\cO(i)\overline{\cO}(i)\Phi_n\right)\Bigr\rangle. \nonumber 
\end{eqnarray}
(We have not been careful about the fact that $f^{(n)\prime}_k(i)$'s actually diverge.) 


\end{document}